\documentclass[twocolumn]{jpsj3-arXiv}
\usepackage[dvips]{color}
\usepackage{txfonts}

\title{Phase diagram and sweep dynamics of a one-dimensional generalized cluster model}

\author{Takumi Ohta\thanks{takumi@yukawa.kyoto-u.ac.jp}, Shu Tanaka, Ippei Danshita, and Keisuke Totsuka}
\inst{
Yukawa Institute for Theoretical Physics, Kyoto University, Kyoto 606-8502, Japan} 

\abst{
We numerically study quantum phase transitions and dynamical properties in the one-dimensional cluster model
with several interactions by using the time-evolving block decimation method for infinite systems and the exact diagonalization.
First, boundaries among several quantum phases of the model are determined from energy gap
and each phase is characterized by order parameters and the entanglement spectrum (ES).
We confirm that in the model with open boundary condition
the degeneracy of the lowest levels in the ES corresponds to that of the ground states.
Then, using the time-dependent Bogoliubov transformation with open boundary condition,
we investigate dynamical properties during an interaction sweep through the critical point
which separates two topological phases involving four-fold degeneracy in the ground state.
After a slow sweep across the critical point, we observe spatially periodic structures
in the string correlation functions and the entanglement entropy.
It is shown that the periodicities stem from the Bogoliubov quasiparticles generated near the critical point.
}

\begin{document}


\maketitle

{\it Introduction--} The exploration of topological phases such as quantum Hall states, topological insulators, and spin liquid phases
has been carried out intensively for three decades.
Topological phases are characterized not by any local order parameters
but by nonlocal order parameters or their emergent edge excitations.
\cite{Nijs-1989,Hatsugai-1991,Kennedy-1992,Hida-1992,Kohmoto-1992,Oshikawa-1992,Yamamoto-1993,Totsuka-1995}
Recently, it was found that entanglement can characterize topological phases.\cite{Kitaev-2006,Levin-2006,Furukawa-2007,Jiang-2012}
Li and Haldane proposed the concept of entanglement spectrum (ES),
which is obtained from the eigenvalues of the reduced density matrix of a subsystem.\cite{Li-2008}
They found that the ES of fractional quantum Hall states is similar to the energy spectrum  of
the low-lying excitations appearing at the edge of the system.
Since then, the ES has been widely used as a tool to study topological phases.
\cite{Pollmann-2010,Hasebe-2011,Cirac-2011,Lou-2011,Fagotti-2011,Tanaka-2012,Qi-2012,Nonne-2013,Cincio-2013,Shinjo-2015}

One of the simplest models which show a variety of topological phases would be the cluster model.\cite{Suzuki-1971,Skrovseth-2009,Smacchia-2011}
The ground state of the model is called the cluster state,\cite{Raussendorf-2001}
which is characterized by the non-local string order parameter.\cite{Smacchia-2011}
In the case of open boundary condition, two zero-energy modes localized at each end of the system exist and four-fold-degenerate ground states result.
The model is interesting also from the quantum-information perspective.
In fact, one-way quantum computation and measurement-based quantum computation using the cluster state
were proposed.\cite{Raussendorf-2001,Raussendorf-2003,Fujii-2013}
In addition, the one-dimensional cluster model is expected to be realized in experiments of cold atoms on a zigzag ladder
by introducing three-spin exchange interaction.\cite{Pachos-2004}

Dynamical properties of systems associated with phase transitions
have been extensively studied for a long time.\cite{Suzuki-1968,Kibble-1976,Zurek-1985}
Dynamics during a parameter sweep across the critical point
especially has been investigated.\cite{Zurek-2005,Mukherjee-2007,Dziarmaga-2010,Polkovnikov-2011,Suzuki-2013,Canovi-2014}
In phases characterized by conventional local order parameters, a universal relation called the Kibble--Zurek scaling\cite{Kibble-1976,Zurek-1985}
is known to hold for the dynamics of defect density.
However, dynamics in topological systems depends on their topological properties and differs from the Kibble--Zurek physics, 
as pointed out in Refs.~\citen{Bermudez-2010,Kells-2014,Hegde-2014}.

In this letter, we focus on topological quantum phase transitions and
dynamical properties under an interaction sweep through a critical point associated with a topological phase transition.
Specifically, we generalize the cluster model in one dimension by adding interaction terms with topological nature,
which give rise to a quantum phase transition between two different cluster states.
Our model would be interesting in the light both of quantum-information science and of competing topological phases.
We map out the ground-state phase diagram of our model by calculating energy gap and
characterize the phases by order parameters and the ES.
In addition, the dynamics during an interaction sweep across the critical point
by calculating energy gap is investigated.
For slow sweep speeds, we observe, after passing the critical point,
periodic structures in the length dependence of string correlation functions and the entanglement entropy (EE).
This breakdown of adiabaticity is due to the topological degeneracy of the initial cluster state.
We conclude that the periodicity is attributed to the Bogoliubov quasiparticles (bogolons).

{\it Model--} The one-dimensional cluster model is defined by
\begin{equation}
\label{eq:clustermodel}
H_{\rm C}=-\sum_{i=1}^{N} J^{XZX}\sigma_i^x\sigma_{i+1}^z\sigma_{i+2}^x,
\end{equation}
where $N$ is the system size and $\sigma_i^{\alpha}$ $(\alpha=x,y,z)$ are the Pauli matrices at site $i$.\cite{Suzuki-1971,Skrovseth-2009,Smacchia-2011}
With open boundary condition, we suppose $\sigma_{N+1}^\alpha=\sigma_{N+2}^\alpha=0$ ($\alpha = x,y,z$).
The three-site interaction in Eq.~(\ref{eq:clustermodel}) is called the cluster interaction or the cluster stabilizer in quantum-information science.
In the ground state of the model dubbed the cluster state, the string order parameter $O_{XZX}=\lim_{L \to \infty} O_{XZX}(L)$ is finite,
where
\begin{equation}
\label{eq:stringcorrelation}
O_{XZX}(L)=(-1)^L \left\langle\sigma_1^x\sigma_2^y\left(\prod_{i=3}^{L-2}\sigma_i^z\right)\sigma_{L-1}^y\sigma_L^x\right\rangle
\end{equation}
is called the string correlation function of length $L$.\cite{Nijs-1989,Kennedy-1992,Smacchia-2011}
A phase characterized by the non-vanishing string order parameter is generally called the cluster (C) phase.
Throughout this letter, we consider a generalization of the cluster model in one dimension to clarify the topological properties.
The {\it generalized cluster model} is defined by adding the Ising interaction $\sigma_{i}^y\sigma_{i+1}^y$ and
another cluster interaction $\sigma_{i}^y\sigma_{i+1}^z\sigma_{i+2}^y$ to the original cluster model:
\begin{equation}
\label{eq:HGC}
H_{\rm GC}=\sum_{i=1}^{N}(-J^{XZX}\sigma_i^x\sigma_{i+1}^z\sigma_{i+2}^x+J^{YY}\sigma_i^y\sigma_{i+1}^y+J^{YZY}\sigma_i^y\sigma_{i+1}^z\sigma_{i+2}^y).
\end{equation}
Note that our model at $J^{YZY}=0$ is the same as the cluster Ising model studied in Ref.~\citen{Smacchia-2011}.
When $J^{YZY}$ is dominant, there appears another topological phase, which we call the dual cluster (C*) phase.
The phase is characterized by the dual string order parameter $O_{YZY}=\lim_{L \to \infty} O_{YZY}(L)$,
where
\begin{equation}
\label{eq:dualstringcorrelation}
O_{YZY}(L)=\left\langle\sigma_1^y\sigma_2^x\left(\prod_{i=3}^{L-2}\sigma_i^z\right)\sigma_{L-1}^x\sigma_L^y\right\rangle
\end{equation}
is called the dual string correlation function of length $L$.

The model (\ref{eq:HGC}) can be solved by using spinless fermion representation.\cite{Lieb-1961}
The original spin model (\ref{eq:HGC}) is transformed into a quadratic Hamiltonian:
\begin{equation}
\label{eq:Hquad}
H=\sum_{i, j=1}^N\left[c_i^{\dagger}A_{ij}c_j+\frac{1}{2}\left(c_i^{\dagger}B_{ij}c_j^{\dagger} +c_iB_{ji}c_j   \right) \right]
\end{equation}
by the Jordan--Wigner transformation:
\begin{equation}
c_i=\prod_{j=1}^{i-1}(-\sigma_j^z)\,\sigma_i^-,\quad c_i^{\dagger}=\prod_{j=1}^{i-1}(-\sigma_j^z)\,\sigma_i^+,
\end{equation}
where $A$ is a real symmetric matrix and $B$ is a real antisymmetric matrix with
$A_{i,i+1}=J^{YY}$, $A_{i,i+2}=J^{XZX}-J^{YZY}$, $B_{i,i+1}=-J^{YY}$, $B_{i,i+2}=J^{XZX}+J^{YZY}$.
In general, quadratic Hamiltonians (\ref{eq:Hquad}) can be diagonalized by the Bogoliubov transformation as
\begin{equation}
\label{eq:HGCBogoliubov}
H=\sum_{\mu=1}^{N} E_{\mu}\left(\eta_{\mu}^{\dagger}\eta_{\mu}-\frac{1}{2}\right),\quad E_{\mu}\geq0.
\end{equation}
The ground state is given by the Bogoliubov vacuum $\left| {\rm vac} \right\rangle$ satisfying $\eta_{\mu} \left| {\rm vac} \right\rangle = 0$ for all $\mu$.

\begin{figure}
\begin{center}
  \includegraphics[scale=1]{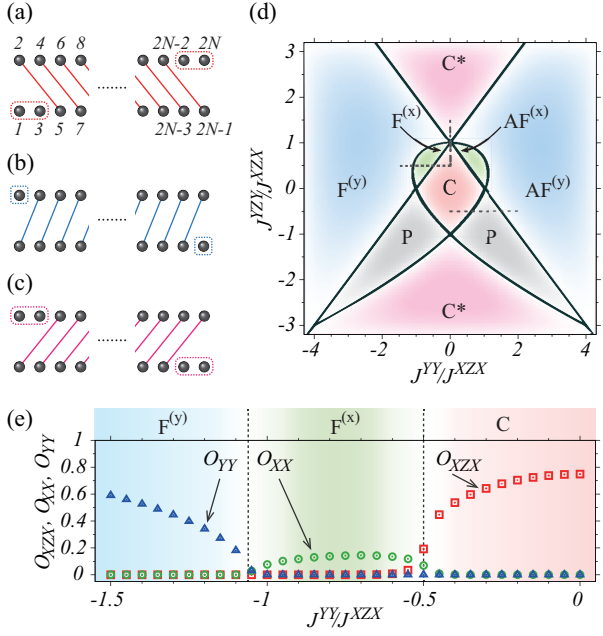} \\
\end{center}
\caption{
(Color Online)
(a)-(c) Schematic representation of the interactions in Eq.~(\ref{eq:HGC}) by the Majorana language.
(a), (b), and (c) depict the first, second, and third terms of the Hamiltonian (\ref{eq:HGC}). 
Non-interacting Majorana fermions enclosed in dotted line appear at the ends of the system.
(d) Phase diagram of the generalized cluster model (\ref{eq:HGC}) for $J^{XZX}>0$.
On the thick solid curves, the excitation gap $\Delta_k$ vanishes at a certain $k$.
C, C*, F, and AF represent cluster, dual cluster, ferromagnetic, and antiferromagnetic phases, respectively.
The P phase cannot be characterized by string and (anti) ferromagnetic order parameters.
The superscript represents the direction of the order.
Each phase is determined by order parameters calculated with the iTEBD.
Along the thin dotted line and the dashed line, we calculated the ES in Fig.~\ref{fig:GCIES} (b), (c).
(e) String order parameter $O_{XZX}$ and ferromagnetic order parameters $O_{XX}, O_{YY}$ in the $x$, $y$-directions
along the thick dotted line in Fig.~\ref{fig:phase} (d) ($J^{YZY}/J^{XZX}=0.5$)
by using the iTEBD with bond dimension $\chi=60$.
}
\label{fig:phase}
\end{figure}

The topological nature of the model (\ref{eq:HGC}) is clearly seen in the Majorana representation.\cite{Kitaev-2001}
The Majorana fermions \{$\bar{c}_i$\} are defined by real and imaginary parts of the two fermion operators \{$c_i$\} and \{$c_i^{\dagger}$\}:
\begin{equation}
\bar{c}_{2i-1}=c_{i}^{\dagger}+c_i,\quad \bar{c}_{2i}=\mathrm{i}\,(c_i-c_{i}^{\dagger}),\quad i=1, 2, \dots, N.
\end{equation}
The standard anticommutation relations of \{$c_i$\} and \{$c_i^{\dagger}$\} translate into
\begin{equation}
\bar{c}_{i}=\bar{c}_{i}^{\dagger},\quad \{\bar{c}_{i},\ \bar{c}_{j}\}=2\delta_{ij}.
\end{equation}
Using the Majorana fermions, the model (\ref{eq:HGC}) is rewritten as
\begin{equation}
\label{eq:HGCmajorana}
H_{\rm GC}=\frac{\mathrm{i}}{2}\bar{c}^{\mathrm{T}}M\bar{c},
\end{equation}
where $\bar{c}=(\bar{c}_{1},\bar{c}_{2},\dots,\bar{c}_{2N})^{\mathrm{T}}$ and $M$ is a real antisymmetric matrix with
$M_{2i-1,2i+2}=-J^{YY}, M_{2i-1,2i+4}=-J^{YZY}, M_{2i,2i+3}=J^{XZX}$.
We can easily read off the number of ground states and that of zero modes from the above Hamiltonian.
Suppose the model with open boundary condition.
When either $J^{XZX}$ or $J^{YZY}$ is dominant,
two zero-energy modes localized at each end of the system exist and result in the four-fold degeneracy in the ground state, as shown in Fig.~\ref{fig:phase} (a), (c).
In contrast, when $J^{YY}$ is dominant, only one zero-energy mode exists and the ground state is two-fold degenerate, as shown in Fig.~\ref{fig:phase} (b).
Thus, the number of zero-modes appearing at the edges of the system characterizes each phase.\cite{Fidkovski-2011}

{\it Phase diagram--} In Fig.~\ref{fig:phase} (d), we show the ground-state phase diagram of the model (\ref{eq:HGC}) and
characterize each phase by order parameters and the ES
by using the method given in Ref.~\citen{Latorre-2004}.
Here we impose the periodic boundary condition.
By performing the Bogoliubov transformation, the model is expressed in the momentum space as
\begin{equation}
H=\sum_{0 \leq k \leq \pi}\Delta_k\,(\eta_k^{\dagger}\eta_k+\eta_{-k}^{\dagger}\eta_{-k}),\quad
\Delta_k=2\sqrt{\epsilon_k^2+\delta_k^2},
\end{equation}
where $\Delta_k\geq0$ is the excitation energy at the wave number $k$ and
\begin{align}
\epsilon_k&=(J^{XZX}-J^{YZY})\cos 2k+J^{YY}\cos k, \\
\delta_k&=(J^{XZX}+J^{YZY})\sin 2k-J^{YY}\sin k.
\end{align}
In the following, we use $J^{XZX}$ as the energy unit and assume that $J^{XZX}$ is positive.
The phase boundaries depicted by the thick solid curves in Fig.~\ref{fig:phase} (d) are determined by the condition $\Delta_k=0$ at a certain $k$.

In order to characterize each phase separated by the curves,
we calculated the order parameters with the time-evolving block decimation method for infinite systems (iTEBD).\cite{Vidal-2007,Orus-2008}
This method is advantageous over the exact diagonalization for the computation of the order parameters with no boundary effect.
Figure~\ref{fig:phase} (e) displays the string order parameter (\ref{eq:stringcorrelation})
and ferromagnetic order parameters $O_{\alpha\alpha}=\lim_{L \to \infty}\left\langle \sigma_1^{\alpha} \sigma_L^{\alpha} \right\rangle$ $(\alpha=x,y)$
along the thick dotted line in Fig.~\ref{fig:phase} (d).
Note that we practically take $L=200$,
at which the order parameters safely converge to their thermodynamic values.
We first consider $J^{YZY}=0$.
When $|J^{YY}/J^{XZX}|$ becomes large, conventional ferromagnetic (F$^{(y)}$) and antiferromagnetic (AF$^{(y)}$) orders in the $y$-direction appear,
which are respectively characterized by the ferromagnetic and antiferromagnetic order parameters
defined by $O_{yy}$ and $\lim_{L \to \infty} (-1)^L \langle \sigma_1^y \sigma_L^y \rangle$.
When $|J^{YY}/J^{XZX}|$ becomes relatively small, on the other hand,
the C phase characterized by the string order parameter (\ref{eq:stringcorrelation}) appears.
Next, various phases appears due to non-zero $J^{YZY}$ are found. 
For positive $J^{YZY}$, the AF and F phases ordered in the $x$-direction (labeled respectively by AF$^{(x)}$ and F$^{(x)}$) appears. 
For negative $J^{YZY}$, on the other hand, we find the paramagnetic (P) phase
which cannot be characterized by the string and the (anti) ferromagnetic order parameters. 
Similar behavior was reported in the cluster-$XY$ model\cite{Montes-2012} and in the transverse Ising model in frustrated lattices\cite{Tanaka-2010}.

\begin{figure}
\begin{center}
  \includegraphics[scale=1]{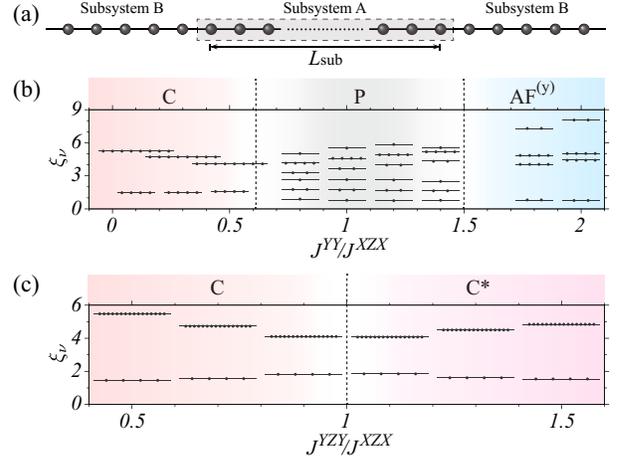} \\
\end{center}
\caption{
(Color Online)
(a) Schematic of subsystems A and B.
(b) ES with $N=503$ and $L_{\rm sub}=249$ for $J^{YZY}/J^{XZX}=-0.5$ indicated by the thin dotted line in Fig.~\ref{fig:phase} (d).
The degeneracy in the lowest levels is four, one (no degeneracy) and two in C, N, and AF phases, respectively.
(c) ES with $N=503$ and $L_{\rm sub}=249$ for $J^{YY}/J^{XZX}=0$ indicated by the dashed line in Fig.~\ref{fig:phase} (d).
The degeneracy in the lowest levels is four in both C and C* phases.
}
\label{fig:GCIES}
\end{figure}

Let us interpret the phase diagram obtained above in terms of entanglement.
Here we suppose open boundary condition and use the exact diagonalization.
We divide the entire system into two subsystems A and B symmetrically around the center
with the length of A being $L_{\rm sub}$ (Fig.~\ref{fig:GCIES} (a)).
We calculate the eigenvalues \{$\lambda_{\nu}$\} of the reduced density matrix $\rho_{\rm A}$ of A defined by tracing out the subsystem B:
$\rho_{\rm A}={\rm Tr}_{\rm B} \rho$,
where $\rho$ is the density matrix of the ground state of the entire system.
The ES is defined as $\xi_{\nu} = -\ln \lambda_{\nu}$.\cite{Li-2008}
First, we focus on the thin dotted line depicted in Fig.~\ref{fig:phase} (d).
As shown in Fig.~\ref{fig:GCIES} (b),
the degree of degeneracy in the lowest value of the ES stays constant in each phase.
We see that its number is four, one, and two for C, P, and AF phases, respectively.
Similarly, focusing on the dashed line depicted in Fig.~\ref{fig:phase} (d)
the degree of degeneracy is four in both C and dual C* phases, as shown in Fig.~\ref{fig:GCIES} (c).
At the critical points, the degeneracy structure disappears.
We also calculate the topological invariant, called the winding number\cite{Anderson-1958}, by using the method explained in Ref.~\citen{Niu-2012}.
We confirm that the winding numbers of C, C*, AF, F, and P phases are two, two, one, one, and zero, respectively,
which correspond to the number of Majorana fermions appearing at the edge of the system (not shown).
From all of these results, the degree of degeneracy is equal to the number of the Majorana zero-modes existing at the ends of the system.
We thus corroborate that for the generalized cluster model (\ref{eq:HGC})
the ES reflects the fictitious edge modes appearing at the position where we cut the system
as in the case of other topological phases studied in previous works.
\cite{Li-2008,Pollmann-2010,Hasebe-2011,Cirac-2011,Lou-2011,Fagotti-2011,Tanaka-2012,Qi-2012,Nonne-2013,Cincio-2013,Shinjo-2015}

\begin{figure}
\begin{center}
  \includegraphics[scale=1]{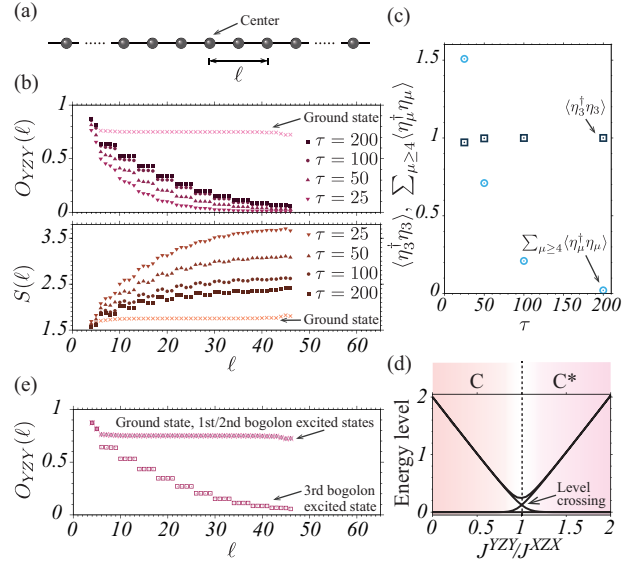} \\
\end{center}
\caption{
(Color Online)
(a) We calculate the dual string correlation function of length $\ell$
and take $\ell$ adjacent sites as the subsystem to calculate the EE.
(b) The length dependence of the dual string correlation function and the EE of final state ($t=\tau$) with $N=101$ and $\tau=$ 25, 50, 100, and 200.
As $\tau$ increases, a quadruple-periodic structure is clearly observed.
(c) The expectation value of the number of bogolons in the final state.
This indicates that only third bogolon is dominant for larger $\tau$.
(d) Energy spectrum of low-lying excited states.
There is a level crossing at the critical point at $J^{YZY}/J^{XZX}=1$.
(e) The dual string correlation function in the states with the bogolon at $J^{YZY}/J^{XZX}=2$.
We can see a quadruple-periodic structure in the length dependence for third bogolon excited state.
The dual string correlation function in the state with the bogolon having zero energy is
the same as that in the Bogoliubov vacuum.
}
\label{fig:Sweep}
\end{figure}

{\it Critical sweep--} Now we turn to the dynamics during an interaction sweep across the critical point between C and C* phases with open boundary condition.
Let us consider the following time-dependent Hamiltonian:
\begin{equation}
H(t)=-J^{XZX}\sum_{i=1}^{N} \sigma_i^x\sigma_{i+1}^z\sigma_{i+2}^x+J(t)\sum_{i=1}^{N} \sigma_i^y\sigma_{i+1}^z\sigma_{i+2}^y,
\end{equation}
where the interaction parameter changes linearly during the sweep time $\tau$ as
\begin{equation}
J(t)/J^{XZX}=2t/\tau,\quad 0\le t \le \tau.
\end{equation}
Here, the eigenenergies in Eq.~(\ref{eq:HGCBogoliubov}) are labelled in ascending order; $E_1 \le E_2 \le \cdots E_N$.
Since $E_1=E_2=0$ except $J^{XZX}=J^{YZY}$,
$\eta_1^\dagger \left| {\rm vac} \right\rangle$, $\eta_2^\dagger \left| {\rm vac} \right\rangle$,
and $\eta_2^\dagger \eta_1^\dagger \left| {\rm vac} \right\rangle$ are the ground states as well as the vacuum state $\left| {\rm vac} \right\rangle$.
The initial state is prepared in the Bogoliubov vacuum corresponding the ground state of the Hamiltonian with $J(t)=0$.
We calculated the length dependence of the dual string correlation function $O_{YZY}(\ell)$ and the EE $S(\ell)$
by using the time-dependent Bogoliubov transformation,\cite{Barouch-1970,Caneva-2007}
where $\ell$ is defined as Fig.~\ref{fig:Sweep} (a).
We define the EE as the von Neumann entropy of $\rho(\ell)$ which is the reduced density matrix of the subsystem
with the length $\ell$, that is, $S(\ell)=-\Tr \rho(\ell)\ln \rho(\ell)$.
Figure~\ref{fig:Sweep} (b) shows the length dependence of the dual string correlation function and the EE
in the final state ($t=\tau$) with $\tau=$ 25, 50, 100, and 200.
For large $\tau$, a quadruple-periodic structure in the length dependence is observed in both quantities.
We calculated the expectation values of the number of bogolons in the final state.
Figure~\ref{fig:Sweep} (c) indicates that only third bogolon is dominant for larger $\tau$.
At $t=\tau/2$, the instantaneous Hamiltonian represents the critical system.
Since the energy levels of low-lying excited states approach the ground states near the critical point as in Fig.~\ref{fig:Sweep} (d),
transitions from the ground state to the excited states would occur for finite $\tau$.
From this fact, one may speculate that the similar periodic structures in the dual string order parameter and the EE
come from the bogolons generated near the critical point.

To substantiate this, we calculated the dual string correlation function for {\it excited} states.
A similar periodic structure is actually observed in the length dependence of the string correlation function
in the excited state with a single bogolon, as shown in Fig.~\ref{fig:Sweep} (e).
This period corresponds to the wave length of the third bogolon.
As mentioned before, the bogolon with zero energy corresponds to the ground-state degeneracy.
Actually, as shown in Fig.~\ref{fig:Sweep} (e), the dual string correlation function in the bogolons with zero energy is
the same as that in the Bogoliubov vacuum.

{\it Conclusion--} We have studied the one-dimensional cluster model with several interactions
to understand the dynamics in the topological phase.
First, we have determined the ground-state topological phase diagram (Fig.~\ref{fig:phase})
and characterized each phase by the order parameters.
In particular, we have found a direct quantum phase transition between two different cluster phases.
We have confirmed that the degeneracy in the lowest levels of the ES in each phase
corresponds to the number of the Majorana fermions existing at the edges of the system (Fig.~\ref{fig:GCIES}).
Second, we have investigated the dynamics during the interaction sweep with finite speeds
across the critical point separating the two cluster states.
The periodicity in the length dependence of the dual string correlation function and the EE has been observed (Fig.~\ref{fig:Sweep}).
We have reproduced similar periodic structure by using the excited states and verified that
the periodicity stems from the bogolons excited when the system is close to the critical point.
It would be interesting to see the results from the viewpoint of topological blocking\cite{Kells-2014,Hegde-2014}.

\begin{acknowledgment}
We thank Koudai Sugimoto and Ryosuke Yoshii for helpful discussions.
This work was supported by JSPS KAKENHI Grant Numbers 25420698 (S.T.), 25800228 (I.D.), 25220711 (I.D.), 24540402 (K.T.).
S.T. is also financially supported by Leave a Nect Co., Ltd. and Discover 21, Inc.
S.T. is the Yukawa Fellow and his work is supported in part by Yukawa Memorial Foundation.
The computations in the present work were performed on super computers at Yukawa Institute for Theoretical Physics, Kyoto University,
and Institute for Solid State Physics, The University of Tokyo.
\end{acknowledgment}


\end{document}